\documentclass[]{article}

\usepackage{times,amsmath,array,epsfig,amsfonts,amssymb,url,graphicx,amsthm,color,epstopdf}
\usepackage[tight,footnotesize]{subfigure}

\usepackage{fixltx2e}
\usepackage[font=footnotesize]{subfig}
\usepackage{multirow}

\usepackage{verbatim}
\title{Fast sequential forensic camera identification}
\author{
{F. P{\'e}rez-Gonz{\'a}lez{\small $~^{\#1}$}, I. Gonz{\'a}lez-Iglesias{\small $~^{\#2}$}, M. Masciopinto{\small $~^{\#3}$}, P. Comesa{\~ n}a{\small $~^{\#4}$}}%
\vspace{1.6mm}\\
\fontsize{10}{10}\selectfont\itshape
$^{\#}$\,Signal Theory and Communications Department, University of Vigo\\
E. E. Telecomunicaci\'on, Campus-Lagoas Marcosende, Vigo 36310, Spain\\
\fontsize{9}{9}\selectfont\ttfamily\upshape
\vspace{1.2mm}\\
\{$^{1}$fperez, $^{2}$igiglesias, $^{3}$mmasciopinto, $^{4}$pcomesan\}@gts.uvigo.es
}

\newcommand{\bt}[1]{\mbox{$\bf #1$}}
\newcommand{\abs}[1]{\lvert #1 \rvert}

\newcommand{\secref}[1]{Sect.~\ref{sec:#1}}
\newcommand{\subsecref}[1]{Sect.~\ref{subsec:#1}}
\newcommand{\figref}[1]{Fig.~\ref{fig:#1}}

\begin{document}
\maketitle

\begin{abstract}
Two sequential camera source identification methods are proposed. Sequential tests implement a log-likelihood ratio test in an incremental way, thus enabling a reliable decision with a minimal number of observations. One of our methods adapts Goljan et al.'s to sequential operation. The second, which offers better performance in terms of error probabilities and average number of test observations, is based on treating the alternative hypothesis as a doubly stochastic model. We also discuss how the standard sequential test can be corrected to account for the event of weak fingerprints. Finally, we validate the goodness of our methods with experiments.      
\end{abstract}

\section{Introduction}\label{sec:intro}

%The origins of the PRNU signal \cite{Holst} \cite{Healey} 

The PhotoResponse NonUniformity (PRNU) is a spatial pattern that acts as fingerprint or unique
identifier of an image camera device. The PRNU is caused by minute imperfections in the image sensor manufacturing process, which remain constant over the life of the device. The PRNU is a variation in pixel responsivity and it arises when the device is illuminated~\cite{Holst}.
Despite the PRNU is generally an extremely weak signal, given enough images samples and using signal processing techniques, it is possible to estimate it and use it as a robust fingerprint~\cite{lukas2006}. This fingerprint can be useful not only for source identification, but also for device linking, fingerprint matching, or forgery detection~\cite{fridrich2009Magazine}. 

% **** mas estado del arte diciendo algunos de forgery otros de fingerprint maching, italianos con bayesian appoach para forgery, temas de alineamiento también hay trabajos en esta linea.... no se.... ejemplos del estado del arte aquí... pero ninguno al final se metió en el propio detector..... ***

% . A reason for trying to obtain the best possible performance for a small number of pixels
% is computational efficiency; indeed, when large numbers of images must be screened,
% it is customary to detect the presence of the PRNU in a small subset of pixels in such
% a way that if the statistic is well below a threshold, the result is considered as ‘no-
% match’. In case that the statistic is above a threshold, the detection is then performed
% over the whole image.
Nowadays, there are huge databases containing thousands of millions of images taken from many different cameras. To analyze whether any of those images was taken from a particular device, the computational cost would be enormous, since each processed image involves operations over more than $10^6$ pixels for a typical image resolution. 

The problem of fast source camera identification has been tackled in previous works~\cite{goljan2013digest, hu2012fast}, always based on the idea of considering a so called \emph{fingerprint digest} instead of the whole fingerprint. However, to the best of our knowledge, there is no available PRNU detection method that works sequentially, i.e, that analyzes small blocks of pixels until enough reliability on the hypothesis test is achieved, thus guaranteeing  that the minimum number of blocks is used to achieve a target accuracy. This is extremely valuable when checking very large amounts of images for the presence of a specific PRNU.  Of particular interest to us is the fast source matching in huge databases used by the police to investigate child pornography and other cybercrime forensic cases. This is the topic of the European Project NIFTY~\cite{URLNifty} under which this work has been carried out.

%**** Motivation by Fernando should be here ****
%**** decimos algo de los chinos que les pedimos el codigo y no contestaron y que proponian un metodo de extraer la %PRNU que supuestamente funcionaba mejor, pero que nosotros no fuimos capaces de que nos funcionara ?? ****

In this paper, we revisit the PRNU estimation and detection problems, and propose a fast algorithm for source camera identification. Firstly, we derive a detector that improves the classical PRNU detection. Next, a sequential algorithm over pseudorandom subsets of pixels is proposed, performing the PRNU detection in a very fast way. Finally, the sequential versions of the detector in~\cite{Goljan2009} and an improved detector are compared in terms of accuracy and average number of iterations of the sequential detector. 

The paper is organized as follows: Sect.~\ref{sec:est_det} reviews PRNU estimation and detection, drawing connections with existing methods, and proposing an improved detector. The fast sequential identification algorithm based on the improved detector is presented in Sect.~\ref{sec:SPRT}. Sect.~\ref{sec:results} shows experimental results on a dataset containing images from several devices, while Sect.~\ref{sec:conclusions} gives our conclusions. 

{\bf Notation}: Vectors are represented in boldface.  The $m$th component of $\bt x$ is denoted as $x_m$. The scalar product of vectors $\bt x$ and $\bt y$ is denoted by $\langle \bt x, \bt y \rangle$, while $\bt x \circ \bt y$ and $||\bt x||$ denote the sample-wise product and the Euclidean norm, respectively.  

\section{Model}
\label{sec:est_det}
We assume that the sensor output at pixel $(i,j)$, $y(i,j)$ can be written as~\cite{Chen2008}
\begin{equation}
\label{eq:model1}
y(i,j) = [1+k(i,j)] x(i,j) + n(i,j),
\end{equation}
where $k(i,j)$ is the (possibly gamma-corrected) PRNU and $n(i,j)$ subsumes a number of noise sources, including dark current, shot noise, read-out noise and quantization noise. As $x(i,j)$ is generally unknown, it is reasonable to obtain an estimate $\hat x(i,j)$ from $y(i,j)$ by applying some denoising procedure and accounting for demosaicing.  In such case, we can write 
\begin{equation}
\label{eq:model2}
y(i,j) = [1+k(i,j)] \cdot [\hat x(i,j)  + r(i,j)] + n(i,j),
\end{equation}
where $r(i,j)$ is the denoising and demosaicing residue. For simplicity, we assume that $n(i,j) \sim  {\mathcal N}(0, \sigma_n^2)$ and $r(i,j) \sim {\mathcal N}(0, \sigma_{i,j}^2)$, where the latter are mutually independent. 

For compactness, we also introduce the {\em shifted PRNU}, which is 1+PRNU, i.e.,  
$\kappa(i,j) \doteq [1+k(i,j)]$.

\subsection{PRNU estimation} \label{sec:PRNUestimation}

From the model in \eqref{eq:model2},  it is possible to formulate the PRNU estimation problem. We recall that in this case we have $L$ available images taken with the same device from which we want to estimate the PRNU at every pixel. Assuming pixel-wise independence, we can solve the estimation problem independently for each pixel. Let then $y_m$ and $\hat x_m$, $m=1, \cdots, L$, denote respectively the observation and the denoised image for an arbitrary pixel of the $m$th available image. Also let $\bt y$, $\hat {\bt x}$ be the vectors formed by stacking the $L$ respective samples for the pixel under analysis. Then, the log-likelihood function becomes 
\begin{equation}
\label{eq:MLE1}
L(\kappa, \hat{\bt x}, \bt y) = - \frac{1}{2} \left[\sum_{m=1}^L \log (2\pi \sigma_{e,m}^2)
+ \sum_{m=1}^L \frac{(\kappa \hat x_m - y_m)^2}{\sigma_{e,m}^2}\right],
\end{equation}
where $\sigma_{e,m}^2 \doteq \kappa^2 \sigma_m^2 + \sigma_n^2$, with $\sigma_m^2$ the variance of the estimation residue in image $m$ (for the $(i,j)$th pixel, $\sigma_m^2=\sigma_{i,j}^2$). 

Taking the derivative of \eqref{eq:MLE1} with respect to $\kappa$ and equating to zero, it is possible to write a (nonlinear) equation that gives the maximum likelihood estimate (MLE) of $\kappa$.

A simpler approach consists in neglecting the first term in \eqref{eq:MLE1}. This gives a minimum weighted MSE solution, namely
\begin{equation}
\hat \kappa =   \arg \min_{\kappa} \sum_{m=1}^L \frac{(\kappa \hat x_m - y_m)^2}{\sigma_{e,m}^2}.
\end{equation}
When $\sigma_m^2= \sigma_r^2$ for all $m=1, \cdots, L$, then taking the derivative with respect to $\kappa$ and setting to zero, we obtain that $\hat \kappa$ must be a solution to the equation
\begin{equation}
\label{eq:quadratic}
\hat \kappa^2 \langle \hat {\bt x}, \bt y \rangle \sigma_r^2 + \left( ||\hat {\bt x}||^2 \sigma_n^2 - ||\bt y||^2 \sigma_r^2 \right) \hat \kappa - \langle \hat {\bt x}, \bt y \rangle \sigma_n^2 =0.
\end{equation}
%where $\langle \cdot, \cdot \rangle$ and $||\cdot||$ denote the inner (scalar) product and Euclidean norm, respectively. 

Assuming that $\sigma_n^2 \gg \sigma_r^2$, the solution to \eqref{eq:quadratic} becomes 
$\hat \kappa = \langle \hat{\bt x}, \bt y \rangle/||\hat {\bt x}||^2$
or equivalently, in terms of the PRNU $\hat k$,
\begin{equation}
\label{eq:kestimation}
\hat k =  \frac{\langle (\bt y - \hat{\bt x}), \hat{\bt x} \rangle}{||{\hat {\bt x}}||^2},
\end{equation}
which in fact resembles Chen et al.'s estimator 
\mbox{$\hat k =  \langle (\bt y - \hat{\bt x}), \bt y \rangle/||{{\bt y}}||^2$} in \cite{Chen2008}, 
as $\hat{\bt x} \approx \bt y$. 

\subsection{PRNU detection}
\label{sec:PRNUdetection}

Once a PRNU estimate is available, it can be used for camera identification purposes. This is in fact a detection problem that can be cast as follows. Given a set of $L$ images which have been taken from the same camera with PRNU ${\bf k}_0$, and a test image $\bt y_t$, both arranged in vector form, we want to decide whether $\bt y_t$ has been taken from that camera or, in other words, if the PRNU $\bt k_0$ is present in $\bt y_t$. 
As customary, we can formulate a binary hypothesis test with the following two hypotheses:
\begin{itemize}
\item[-] $H_0$: Image $\bt y_t$ does not contain the PRNU $\bt k_0$, 
\item[-] $H_1$: Image $\bt y_t$ contains the PRNU $\bt k_0$.
\end{itemize}
We assume the existence of an unbiased estimate $\hat {\bt k}$ of $\bt k_0$ obtained using the method proposed in the previous section, and we denote by $\hat {\bt x}_t$ the image vector obtained from $\bt y_t$ after denoising and demosaicing. We have derived the corresponding distributions under each hypothesis; this allows us to write Neyman-Pearson's generic detector for known $\bt k_0$, and later replace the needed statistics by their estimates, as in the {\em Generalized Likelihood Ratio Test} (GLRT). 

Thus, when $H_1$ holds, we can see that the difference $y_{t}(i,j) - \hat x_{t}(i,j)$ for the $(i,j)$th pixel is Gaussian with mean $k_{0}(i,j) \cdot \hat x_{t}(i,j)$ and variance 
\begin{equation}
\label{eq:varH1}
\sigma^2_{H} = [1+ k_{0}(i,j)]^2 \sigma_r^2 + \sigma_n^2 \approx \sigma_r^2 + \sigma_n^2.
\end{equation}
When $H_0$ holds, the PRNU must be treated as unknown. Modeling it as a zero-mean random variable, it follows that  $y_{t}(i,j) - \hat x_{t}(i,j)$  is approximately Gaussian with zero mean and variance approximately $\sigma_{H}^2$, because the influence of the variance of the PRNU in the total variance is negligible. From this, the likelihood-ratio test can be written as
\begin{equation}
\label{eq:NPtest}
\frac{\langle (\bt y_t - \hat {\bt x}_t) , {\bt k}_0 \circ \hat{\bt x}_t \rangle}{\sigma_H^2} - \frac{||{\bt k}_0 \circ \hat {\bt x}_t||^2}{2 \sigma_H^2} \ {\substack{H_1\\>\\<\\H_0}} \ \eta  
\end{equation}
for some threshold $\eta$ that is chosen so as to produce the desired probability of false positive. 

The implementation \eqref{eq:NPtest} faces two practical problems: 1) The true PRNU ${\bt k}_0$ is unknown; 2) $\sigma_H^2$ is unknown. To overcome the first problem, one may think of substituting $\langle (\bt y_t - \hat {\bt x}_t) , {\bt k}_0 \circ \hat{\bt x}_t \rangle$ by  $\langle (\bt y_t - \hat {\bt x}_t) , \hat {\bt k} \circ \hat{\bt x}_t \rangle$ after noticing that 
\begin{equation}
\langle (\bt y_t - \hat {\bt x}_t) , {\bt k}_0 \circ \hat{\bt x}_t \rangle = E\left\{\langle (\bt y_t - \hat {\bt x}_t) , \hat{\bt k} \circ \hat{\bt x}_t \rangle \right\}.
\end{equation}
However, the results obtained by following this approach are rather disappointing in practice because the variance of the estimation error in $\hat{\bt k}$ significantly affects the computation of the second summand in \eqref{eq:NPtest}.  

Focusing on the first term of \eqref{eq:NPtest} produces one (generally, non-sufficient) statistic that is very similar to which has been proposed by Goljan {\it et al.}~\cite{Goljan2009}:\footnote{Goljan et al. use $\bt y_t$ instead of $\hat{\bt x}_t$ in the second term of the scalar product.}
\begin{equation}
\label{eq:scprod}
u \doteq \langle \bt (\bt y_t - \hat {\bt x}_t), \hat{\bt k} \circ \hat{\bt x}_t \rangle,
\end{equation}
and, as we argue above, the means for the respective hypotheses are
\begin{eqnarray}
E\{u|H_0\}=0; \ \ \  E\{u|H_1\}=E\{||\bt k_0 \circ \bt x_t||^2\}.
\end{eqnarray}

Let us define the shift operator $\Delta_{(q_1,q_2)}$ that applied to a vector $\bt x$ representing an image, outputs the vector corresponding to a right circular shift of $(q_1,q_2)$ pixels of such image. Then, following~\cite{Goljan2009}, an estimate of the variance of statistic $u$ can be obtained as,  
\begin{equation}
\label{JF_W}
%\hat {\sigma}_u^2 = \frac{1}{M}||\hat {\bt k} \circ \hat{\bt x}_t||^2 ||(\bt y_t - \hat {\bt x}_t)||^2
\hat {\sigma}_u^2 = \frac{1}{M-|A|} \sum_{(q_1,q_2) \not\in A} \langle \Delta_{(q_1,q_2)} (\bt y_t - \hat {\bt x}_t) , \hat {\bt k} \circ \hat{\bt x}_t \rangle^2,
\end{equation}
where $M$ is the number of available pixels, $A$ is an {\em exclusion set} defined as those $(q_1,q_2)$ in a neighborhood (w. r. t. circular shifts) of the origin $(0,0)$, and $|A|$ denotes its cardinality. 

Since $E\{u|H_1\}$ is difficult to obtain accurately, it may be reasonable to assume that under $H_1$ the statistic $u$ has a positive but unknown mean.  Then, from Karlin-Rubin theorem~\cite{karlin1956theory}, the test
\begin{equation}
\label{eq:KR}
u' \ {\substack{H_1\\>\\<\\H_0}} \ \eta_2, 
\end{equation}
where $u' \doteq u/\hat{\sigma}_u$, is the uniformly most powerful test for a given probability of false positive $P_F$. However, notice that even though the test threshold $\eta_2$ can be set since $P_F$ is computable, we cannot find the detection probability $P_D$ as the mean under $H_1$ is unknown. This has important implications for the tests discussed in \secref{SPRT}. 
Obviously, better performance would be expected if $\mu_{u,1} \doteq E\{u'|H_1\}$ were known. 

\subsection{Improved detector}
\label{sec:improved}
Although $\mu_{u,1}$ is not known, we have found that it can be modeled as a normal random variable whose parameters depend on the statistic
\begin{equation}
\label{eq:v}
v \doteq || \hat{\bt k} \circ \hat{\bt x}_t||^2/\hat \sigma_u,
\end{equation}
so we will denote by $\mu(v)$ and $\sigma^2(v)$ the mean and variance of $\mu_{u,1}$,  respectively. Thus, $\mu_{u,1} \sim {\mathcal N} (\mu(v), \sigma^2(v))$. 
Notice that the statistic $v$ resembles the second term in \eqref{eq:NPtest} 
but obtained from computable quantities. 
However, the laws $\mu(v)$ and $\sigma^2(v)$ are device-dependent, so they must be learned during the PRNU extraction phase, which can be done concurrently with the estimation of $\bt k_0$. 
Furthermore, although the distribution of $u'$ under $H_0$ can be modeled by a Gaussian, a slight improvement is afforded by employing a zero-mean generalized Gaussian distribution with scale parameter $\alpha_0$ and shape parameter $c_0$, which can be reliably estimated using images from different cameras~\cite{GGD_DCT} (see Sect.~\ref{subsec:TrainingPhase}).

With all these considerations, the test becomes
\begin{equation}
\label{OurDetector}
\left( \frac{\abs{u'}}{\alpha_0} \right)^{c_0}- \frac{(u'-\mu(v))^2}{2\sigma^2(v)} \ {\substack{H_1\\>\\<\\H_0}} \ \eta_3.
\end{equation}

\subsection{Model training} 
\label{subsec:TrainingPhase}

At this point, it is necessary to estimate the model parameters for both hypotheses. For $H_1$, $\mu(v)$ and $\sigma^2(v)$ must be estimated for the target camera, while for $H_0$ the 
parameters $\alpha_0$ and $c_0$ are estimated from the universe of available images.

%The images used in the PRNU extraction give profitable information about the relation between $u'$ and $v$. 
As mentioned, the estimation of the mean and variance of $\mu_{u,1}$ is done concurrently with the estimation of $\hat{\bt k}$. Let $\bt y_{tr}$ be one of the $L$ available images for training. Then, $\hat{\bt k}$ is estimated from the remaining $L-1$ images, and pairs of $(u',v)$ values are obtained from $\bt y_{tr}$ and $\hat{\bt k}$ by taking subsets of pixels with the same size as that used in the hypothesis test. This process is repeated for each of the $L$ images in the training set to produce a collection of $(u',v)$ pairs that is used to estimate $\mu(v)$ and $\sigma^2(v)$. This is done by binning the values of $v$ and for each bin calculating the mean and variance of the corresponding set of $u'$ values.   

\begin{figure}[!t]
\centering
\includegraphics[width=3.5in]{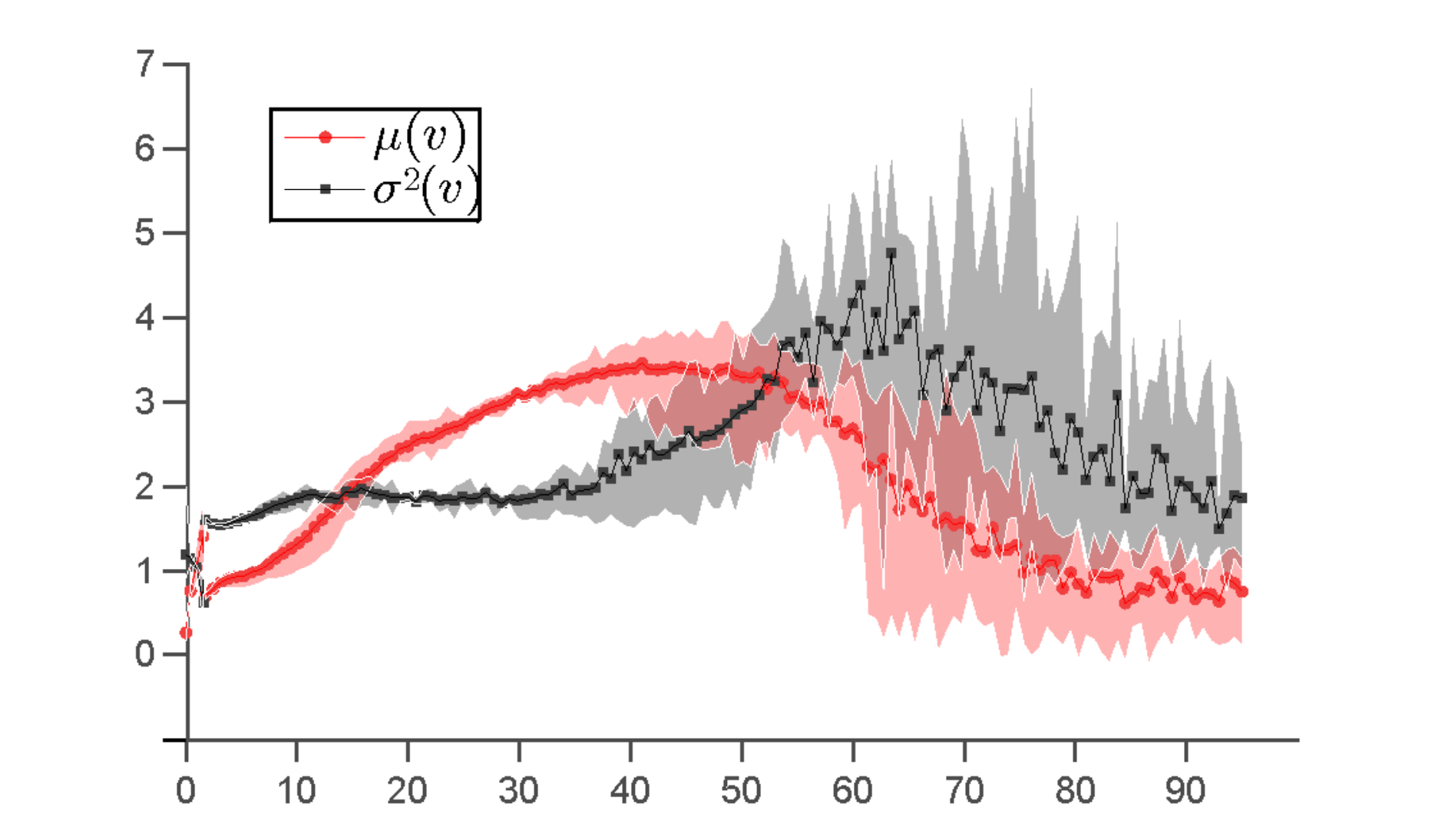}
\caption{Parameters $\mu(v)$ and $\sigma^2(v)$ learned from $L = 50$ images during the training phase for camera Nikon D60.}
\label{fig:training}
\end{figure}

For illustration purposes, Fig.~\ref{fig:training} shows an example of the laws $\mu(v)$ and $\sigma^2(v)$ obtained by applying the explained procedure. %These training phase is learned from the $L$ images of the camera device under investigation (hypothesis $H_1$). 
%The color bands contain 95\% of the observations, showing that the variance of the estimators increases with $v$ as there are less $u'$ values in the corresponding bins. 
The color bands represent the range of values for $5$ different trainings (each training with $L=50$ randomly selected images) over the same device. 
%The variance of the estimators increases with $v$ as there are less $u'$ values in the corresponding bins.

%As it can be observed, for larger values of $v$ Figure~\ref{fig:training} also shows that, the more the value of the %pixels is ($v$), the higher the presence of the PRNU will be ($\mu(v)$), as well as it is mentioned in \cite{Chen2008}.

When the subsets of pixels used for training do not have the same size as for the hypothesis test, the scaling factor $\sqrt{M_{t}/M_{tr}}$ must be applied on $v$, $\mu(v)$ and $\sigma^2(v)$, where $M_{tr}$ and $M_t$ stand for the number of available training and testing pixels, respectively. We note, however, that if the sizes are significantly different, the correction may yield unsatisfactory results. 
%On the other hand, the $\alpha_0$ and $c_0$ parameters of the hypothesis $H_0$ are obtained by estimating the %fingerprint $\hat {\bm k}$ of several cameras, and comparing each fingerprint against images from other different %cameras.

\section{Sequential test for fast PRNU detection} \label{sec:SPRT}

The Sequential Probability Ratio Test (SPRT) was proposed by A. Wald in~\cite{Waldarticle}. On a hypothesis testing problem, the main purpose of the SPRT is minimize the expected number of observations to achieve error probabilities less than a pre-fixed target probabilities of misdetection ($P_M^*$) and false positive ($P_{F}^*$). 

Let $x_1, x_2, \ldots, x_n$ be i.i.d. observations, with $n$ its number, and $f(x_j|H_i)$ the probability density function (pdf) of the $j$th observation conditioned on the $i$th hypothesis ($i = 0, 1$). In a SPRT the likelihood ratio is compared with two thresholds. If
\begin{equation}
B < \prod_{j=1}^{n} \frac{f(x_j|H_1)}{f(x_j|H_0)} < A, \label{notdecide} 
\end{equation}
the test continues and another observation (a.o.) is taken. If
\begin{equation}
 \prod_{j=1}^{n} \frac{f(x_j|H_1)}{f(x_j|H_0)} \geq A, \label{decideH1}
\end{equation}
the test accepts the alternative hypothesis ($H_1$). If
\begin{equation}
\prod_{j=1}^{n} \frac{f(x_j|H_1)}{f(x_j|H_0)} \leq B, \label{decideH0}
\end{equation}
the test accepts the null hypothesis ($H_0$).

The thresholds are chosen so as to control the error probabilities on each hypothesis. Following Wald's approximation~\cite{Waldarticle}, the relations between the maximum permissible errors ($P_M^*$ and $P_F^*$) and the thresholds are
\begin{equation}
\label{AB_thresholds}
A  \leq  \frac{1-P_M^*}{P_{F}^*},\ \ B \geq \frac{P_M^*}{1-P_{F}^*}, 
\end{equation}
where the equality is usually a good choice in practice. 

Therefore, the hypothesis testing problem discussed in \secref{PRNUdetection} and, hence, our proposed detector in \eqref{OurDetector}, can be transformed into a SPRT by taking logarithms in (\ref{notdecide}-\ref{decideH0}). %However, in order to determine the values of the thresholds for the SPRT, it is necessary to derive the log-likelihood %ratio for the detector proposed in \secref{improved} obtaining
%\begin{eqnarray}
%&& \left( \frac{\abs{u'}}{\alpha_0} \right)^{c_0} - \frac{(u'-\mu(v))^2}{2\sigma^2(v)} - log \left( %c_0\sqrt{2\pi\sigma^2(v)} \right) \nonumber \\
%&&{}+log \left(2\alpha_0\Gamma(1/c_0) \right).
%\end{eqnarray}
%\begin{equation}
% \left( \frac{\abs{u'}}{\alpha_0} \right)^{c_0} - \frac{(u'-\mu(v))^2}{2\sigma^2(v)} - log\left( %c_0\sqrt{2\pi\sigma^2(v)} \right) +log\left( 2\alpha_0\Gamma\left(\frac{1}{c_0}\right) \right),
%% \end{equation}
Then, the resulting SPRT is
% \begin{eqnarray}
% &&\mbox{If } 2 \cdot \log(A) < \sum_{j=1}^n D_j  < 2 \cdot log(B) \mbox{, }  \nonumber \\ 
% &&\mbox{   take another observation.}  \nonumber \\
% &&\mbox{If } \sum_{j=1}^n D_j \geq 2 \cdot log(A)\mbox{, decide $H_1$.} \nonumber \\
% &&\mbox{If } \sum_{j=1}^n D_j \leq 2 \cdot log(B)\mbox{, decide $H_0$.} \label{OurDetectorSPRT}
% \end{eqnarray}
% where the detector has the expression
% \begin{equation}
% D_j = \frac{(u'_j)^2}{\sigma_0^2} - \frac{(u'_j-\mu(v_j))^2}{\sigma^2(v_j)} + 2 \cdot log \left( \frac{\sigma_0}{\sigma(v_j)} \right). 
% \end{equation}
\begin{equation}
\eta_B \ {\substack{H_0\\>\\<\\a.o.}} \ \sum_{j=1}^n D_j  \ {\substack{H_1\\>\\<\\a.o.}} \  \eta_A , 
\label{OurDetectorSPRT}
\end{equation}
where the thresholds are $\eta_A = \log(A)-n\cdot \log(2\alpha_0\Gamma(1/c_0))$ and $\eta_B = \log(B)-n\cdot \log(2\alpha_0\Gamma(1/c_0))$, and
\begin{equation}
D_j \doteq \left( \frac{\abs{u'_j}}{\alpha_0} \right)^{c_0} - \frac{(u'_j-\mu(v_j))^2}{2\sigma^2(v_j)} - \log\left(c_0\sqrt{2\pi\sigma^2(v_j)} \right).
\end{equation}
%stands for the obtained sequential detector.

\begin{figure}[!t]
\centering
\includegraphics[width=3.5in]{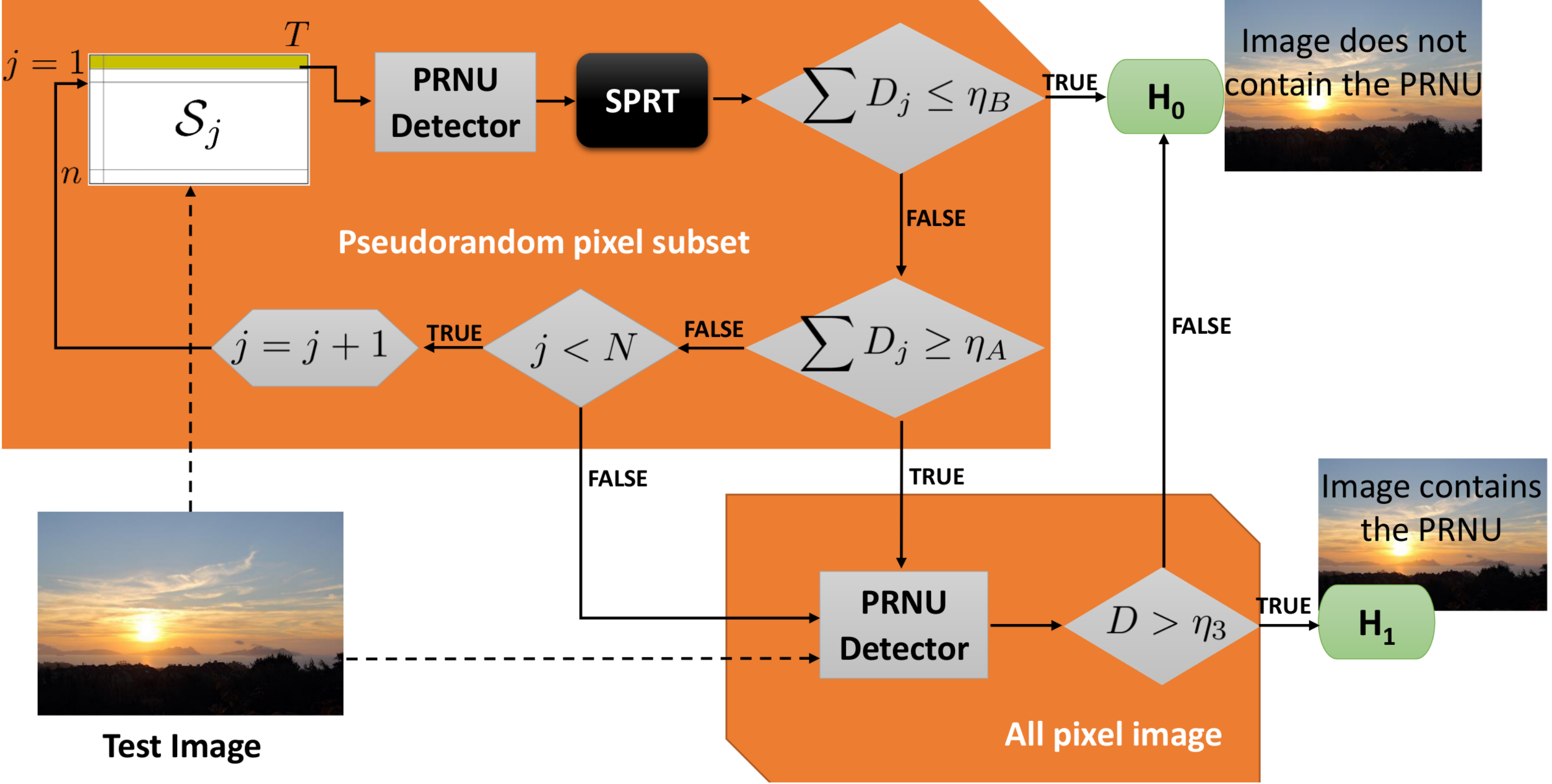}
\caption{Sequential test implementation.}
\label{fig:Diagram}
\end{figure}

Figure~\ref{fig:Diagram} summarizes the proposed algorithm for fast source camera identification. Given a test image $\bt y_t$, the pixels are pseudorandomly assigned to subsets ${\mathcal S}_j$, $j=1, \cdots, n$, with $T$ pixels each. The $j$th observation $(u'_j,v_j)$ is computed by using in \eqref{eq:scprod} and \eqref{eq:v} only those pixels in  subset ${\mathcal S}_j$. A maximum number of observations $N$ is set for the SPRT; if $n$ reaches that value without a decision being taken, the entire image is analyzed with a non-sequential test. In addition, those images classified as $H_1$ by the SPRT are retested using the whole image, in order to achieve the minimum possible error probabilities. 

In setting the thresholds $A$ and $B$, we notice that we aim at achieving a very small probability of misdetection $P_M$, whereas we do not care as much about $P_F$, because the subsequent full-image test will discard most false positives.

%\cite{siegmundsequential}

%\begin{comment}
\subsection{Fixed-parameter PRNU Sequential Test}
\label{AdaptingSPRT}
%A SPRT based on the classical PRNU detection method proposed by Goljan et al. in~\cite{Goljan2009} is derived, in order to compare it with the method proposed in \secref{improved}.

Here we propose a SPRT based on the well-known detection method by Goljan et al. in~\cite{Goljan2009}. Such method 
sets the detection threshold on the basis of a target $P_F$, but entirely disregards hypothesis $H_1$ and, consequently, $P_M$. See \eqref{eq:KR}. However, knowledge of $P_M$ is necessary to implement a SPRT as is apparent from \eqref{AB_thresholds}. 

In contrast,  our detector in \eqref{OurDetector} derived in Sect.~\ref{sec:improved} overcomes this issue, as now the distribution of $u'$ under $H_1$ is well defined. On the other hand, as we noticed in Sect.~\ref{subsec:TrainingPhase}, in some cases a deficient training phase may lead to bad estimates of $\mu(v)$ and $\sigma^2(v)$, which would have a large impact on the detection performance.  To propose a feasible solution to those cases and at the same time quantify how much is gained by learning $\mu(v)$ and $\sigma^2(v)$, we have also studied the performance of our detector in  \eqref{OurDetector}
when $\mu(v)$ and $\sigma^2(v)$ are assumed to be independent of $v$. The results are reported in  Sect.~\ref{sec:results}.

\subsection{Dealing with weak PRNUs} \label{subsec:missmodeling}

\begin{figure}[!t]
\centering
\includegraphics[width=2.5in]{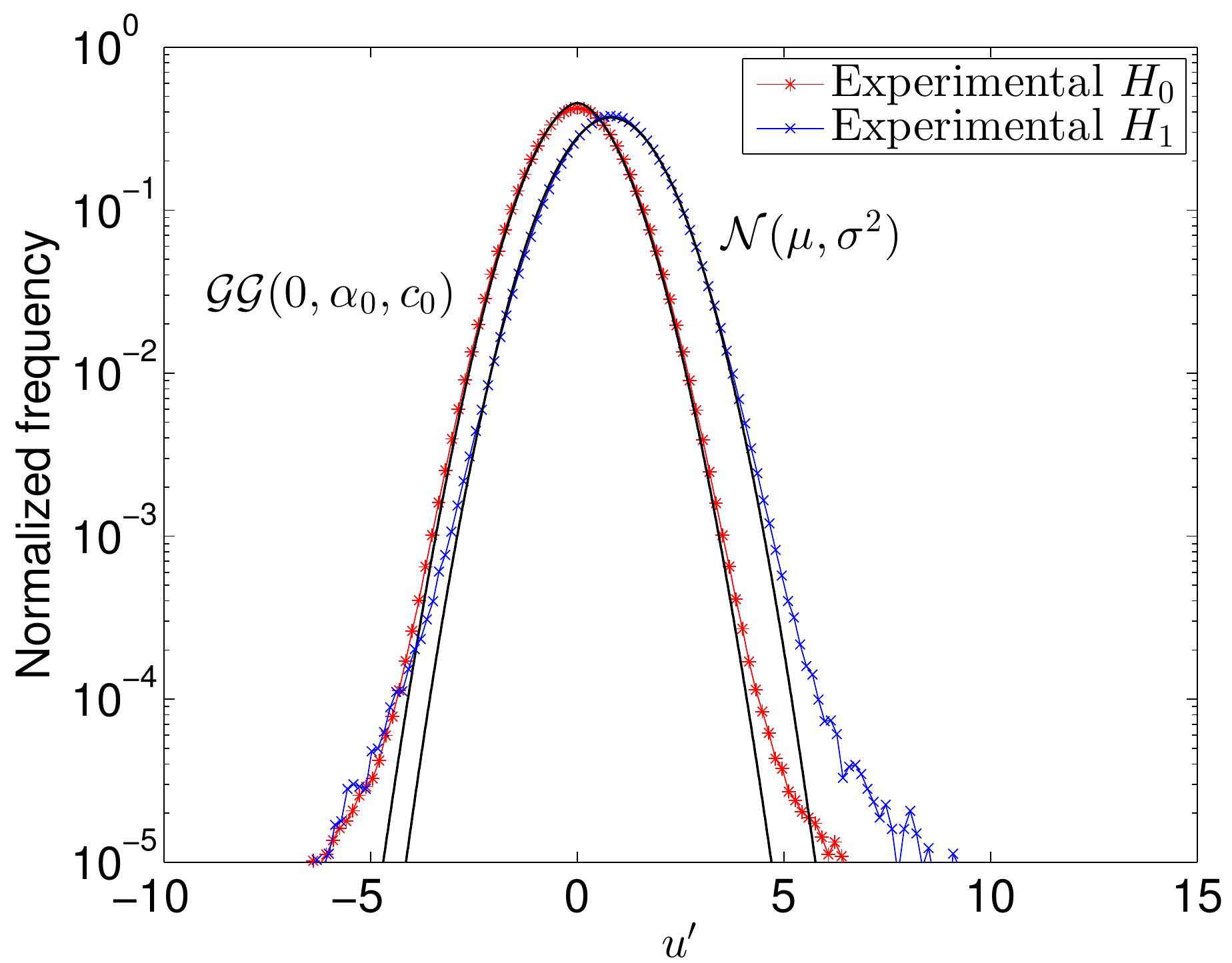}
\caption{Log-plot of experimental and theoretical distributions of $u'$ under both hypotheses for $1024$ random pixels. Camera Nikon D60 is used for the hypothesis $H_1$, resulting on \mbox{$\mu = 0.81$} and \mbox{$\sigma^2 = 1.17$}. The parameters for $H_0$ are $\alpha_0 = 1.24$ and \mbox{$c_0 = 1.78$}.}
\label{fig:Histograms_H0_H1}
\end{figure}

Although the model used in \secref{improved} fits quite well the statistical distributions for both hypotheses,  there are cases where the PRNU is weak due to the contents of the image~\cite{Chen2008}. As discussed in Sect.~\ref{subsec:TrainingPhase}, the mean $\mu(v_j)$ may vary significantly among different observed subsets ${\mathcal S}_j$. For very dark images $v \approx 0$ and $\mu(v)$ may be very close to zero, implying that both hypotheses are barely distinguishable. Another problematic case occurs when an image presents many white or saturated pixels at any color channel, since saturated pixels are PRNU-free. Despite the model fits quite well $u'$ under hypothesis $H_1$, we have noticed experimentally that some images contain most of the observations $u'_j$ on the left tail of $H_1$, to the point that those images are wrongly classified as $H_0$ after the SPRT has processed a few initial observations. See Fig.~\ref{fig:Histograms_H0_H1} for an illustration of the overlap between the left tails of $H_0$ and $H_1$, with also their matching pdf's for fixed parameters, i.e., independent of $v$. 

In order to address those outliers, we introduce a probability $p \in \left [ 0, 1 \right )$ that the observations $u'_j|H_1$ follow the distribution corresponding to $H_0$ (i.e., the probability that the observations for $H_1$ do not contain the PRNU). We next show that when $p >0$ the SPRT requires that the thresholds $A$ and $B$ be modified as $P_M$ will change ($P_F$ does not, as $H_0$ remains the same). If $P'_M$ denotes the new misdetection probability, it is easy to show that  $P'_M = p \cdot (1-P_F) + (1-p) \cdot P_M$. 

Therefore, to achieve a certain target misdetection probability $P^{'*}_M$, the thresholds must be recomputed by substituting 
\begin{equation}
P_M^* = \frac{P^{'*}_M - p \cdot (1-P_F^*)}{1-p} \label{eq:P_M}
\end{equation}
into the expressions in \eqref{AB_thresholds}. An important remark is that the presence of outliers imposes a bound on the achievable detection probability $P^*_D=(1-P^*_M)$, namely, $P_D^* \leq 1 - p \cdot (1-P_F^*)$.

%establishes an upper limit for $P_D^*$ for a given $p$ and $P_F^*$. For practical uses, where the aim is to achieve %the target probabilities $P_D^*$ and $P_F^*$, the following upper limit of $p$ can be obtained from \eqref{Pd_max}
%\begin{equation}
%\label{p_max}
%p < \frac{1-P_D^*}{1-P_F^*}.
%\end{equation}

Thus, in practice, the maximum achievable $P_D$ may be below the target. If such is the case, then it is necessary to give up on the target $P_F$, i.e., a larger value will be achieved. This has a limited impact in practice because we remind that all the positives from the SPRT are later subject to a full-image test which will discard most of those false positives. The increase of $P_D$ can be achieved by multiplying both $A$ and $B$ by a factor $\beta \leq 1$. 

\subsection{Fast variance estimation}
\label{sec:fve}
The variance estimator $\hat \sigma^2_u$ in \eqref{JF_W} is taken from~\cite{Goljan2009}, which is quite time-consuming as it contains two nested sums (corresponding to the scalar product and the averaging over the spatial shifts). Since one of the aims of the methods proposed in this paper is to reduce the detection time, we propose the following 
simpler estimator
\begin{equation}
\label{OurW}
\hat {\sigma}_u^{'2} = \frac{1}{M}||\hat {\bt k} \circ \hat{\bt x}_t||^2 ||(\bt y_t - \hat {\bt x}_t)||^2.
\end{equation}

\begin{figure}[!t]
\centering
\includegraphics[width=2.5in]{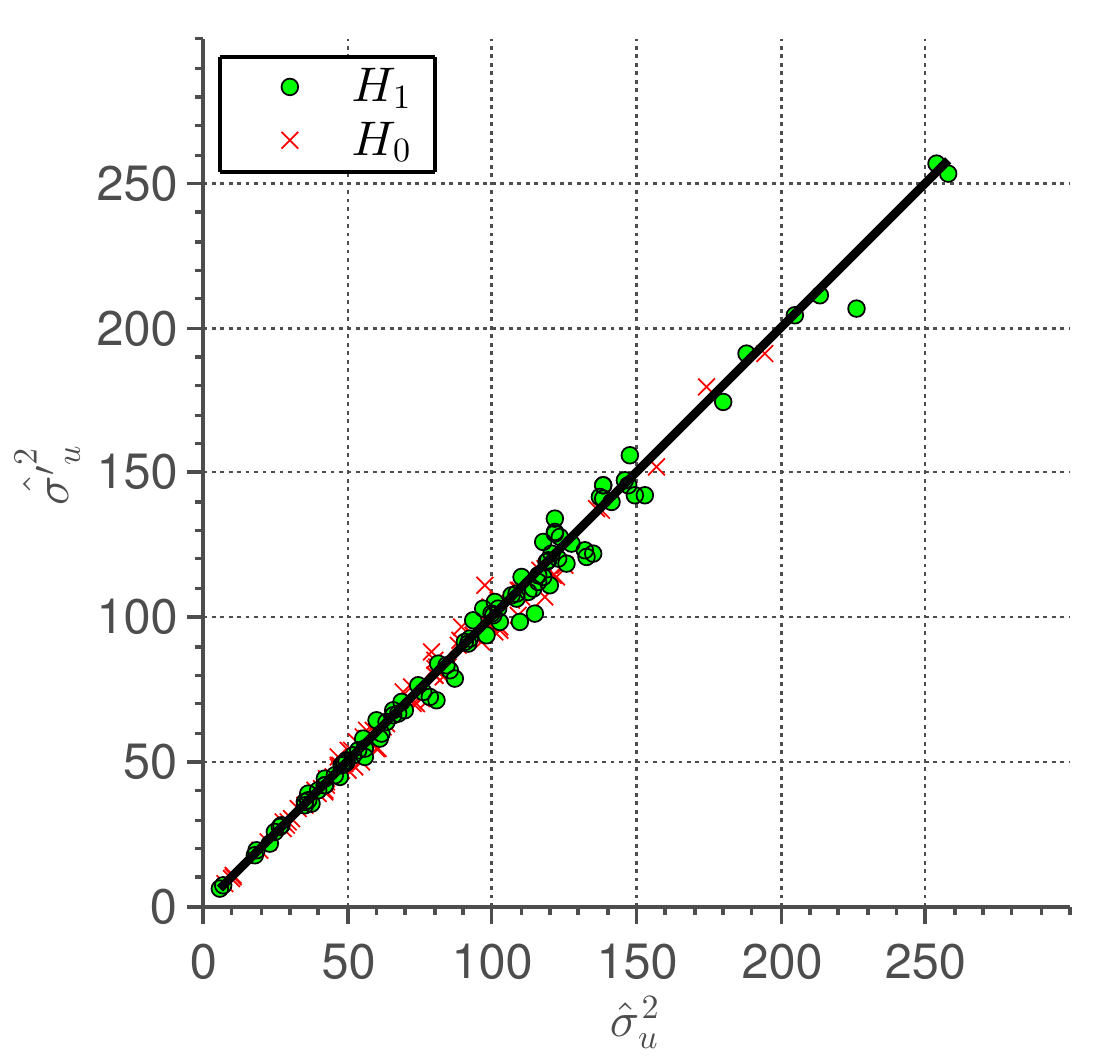}
\caption{Values of $\hat{\sigma}_u^2$ versus $\hat{\sigma'}_u^2$ for $100$ samples from each hypothesis. Each sample is from a $32 \times 32$ block-size for $147$ images taken with Nikon D60.}
\label{fig:w}
\end{figure}

Figure \ref{fig:w} shows that there is a very little difference between $\hat{\sigma}_u^2$ and $\hat{\sigma}_u^{'2}$ under both hypotheses. In fact, in terms of camera identification performance, there is no significant difference between both variance estimation methods, since the Areas Under Curve (AUCs) for the respective Receiver Operating Characteristic (ROC) curves are practically identical (differences show up in the $4$th significant digit).
%, as \figref{w_ROC} shows.

In terms of computing time, if $M$ is the total number of pixels in the image and $|A|$ is the size of the exclusion set, the estimator in \eqref{OurW} is $M-|A| \approx M$ times faster, which even for small-sized images results in enormous savings. 

%Let $t_M$ be the time to compute the expression \mbox{$||\hat {\bt k} \circ \hat{\bt x}_t||^2 ||(\bt y_t - \hat {\bt %x}_t)||^2$}, which depends of the number of pixels $M$, measured in units of time (UT). Then, to compute %\eqref{JF_W},  $M \cdot t_M$ UT are necessary, since it implies the cross-correlation over all possible shifts of the %pixels of the image under test $\bm y_t$. Therefore, the computation of \eqref{OurW} is $M$ times faster than to %compute \eqref{JF_W}. 

%Hence, for practical purposes where a group of $N_t$ images from a database are analyzed in order to link them with %the camera under test, the saved time is
%\begin{equation}
%\label{savedtime}
%  N_t\cdot (M-1) \cdot t_M [UT].
%\end{equation}

%Notice that, in the expression \eqref{savedtime} either $M$ or $N_t$ can be, at least, around $10^3$, and depending %of the scenario, the number of images $N_t$ can be very large.
%Hence, for practical purposes where the computational time is important, every image $\bt y_t$ that we want to %analyze save $(M-1) \cdot t_M$ UTs by using \eqref{OurW}. 
%For instance, if the aim is to apply the detection algorithm on a huge database, containing thousands or millions of %images, the time saved by computing \eqref{OurW} instead of \eqref{JF_W} is very significant.  

%In the case of SPRT, the simplification \eqref{OurW} on the evaluation of $\hat {\sigma}_u^2$ for each $\bm y_t$ is even more noticeable, since for every observation  

\section{Experimental results}
\label{sec:results}
The image database for the experimental results is a collection of pictures from different sources. The database is composed of TIFF images coming from our own cameras, the Dresden image database~\cite{Gloe}, and the Raise database~\cite{Raise}. Some camera models include more than one device as Table~\ref{tab:database} shows. The table specifies the number of images from each device as well as some camera model characteristics.
\begin{table}[!h] 
%% increase table row spacing, adjust to taste
\renewcommand{\arraystretch}{1.3}
% if using array.sty, it might be a good idea to tweak the value of
% \extrarowheight as needed to properly center the text within the cells
\caption{Cameras used in SPRT experiments with characteristics, number of devices for each model, number of images for each device and database source.}
\label{tab:database}
\centering
%% Some packages, such as MDW tools, offer better commands for making tables
%% than the plain LaTeX2e tabular which is used here.
\scalebox{0.8}{
\begin{tabular}{c c c c c c}
\hline
\hline
\centering
Camera Model & Sensor & \multicolumn{1}{m{1.5cm}}{\centering  Native \\ resolution} & Devices & \multicolumn{1}{m{2cm}}{\centering Number of \\ images} & Database \\
\hline
Canon 600D & 22.3x14.9mm CMOS & 5184x3456 & 1 & 241 & Own\\ 
Canon1100D & 23.2x14.7mm CMOS & 4272x2848 & 3 & 316/122/216 & Own\\ 
Nikon D60 & 23.6x15.8mm CCD & 3872x2592 & 1 & 197 & Own \\
Nikon D70 & 23.7x15.6mm CCD & 3008x2000 & 2 & 43/43 & Dresden \\
Nikon D70S & 23.7x15.6mm CCD & 3008x2000 & 2 & 43/47 & Dresden \\
Nikon D90 & 23.6x15.8mm CMOS & 4288x2848 & 1 & 250 & Raise\\
Nikon D200 & 23.6x15.8mm CCD & 3872x2592 & 2 & 48/43 & Dresden\\
Nikon D3000 & 23.6x15.8mm CCD & 3872x2592 & 1 & 230 & Own\\
Nikon D3200 & 23.2x15.4mm CMOS & 6016x4000 & 1 & 250 & Own\\
Nikon D5100 & 23.6x15.6mm CMOS & 4928x3264 & 1 & 250 & Own\\
Nikon D7000 & 23.6x15.6mm CMOS & 4928x3264 & 1 & 250 & Raise\\
\hline
\hline
\end{tabular}
}
\end{table}
 
For PRNU extraction, 
$L = 50$ images are randomly selected for those devices in Table~\ref{tab:database} with more than $50$ images available. The PRNU is extracted as described in \secref{PRNUestimation}. 
The experimental results sequentially pick as hypothesis $H_1$ each
device with more than $50$ images, and $H_0$ all the database images from the remaining devices in Table~\ref{tab:database}.  

In order to make the results independent of the specific choice of $L$ images and also to increase the number of test images corresponding to $H_1$, the reported results are the average of $5$ different random selections of the $L$ images, using the remaining images of each selection to test $H_1$.

In all the experiments, the denoised images $\hat{\bt x}_t$  are obtained using the same filter as in~\cite{Chen2008}. In addition, the estimated PRNU in \eqref{eq:kestimation} is postprocessed to remove the unwanted artifacts discussed in~\cite{Chen2008}; this postprocessing includes mean-subtraction and Wiener filtering in the Fourier domain. 

\begin{figure}[!t]
\centering
\includegraphics[width=3.5in]{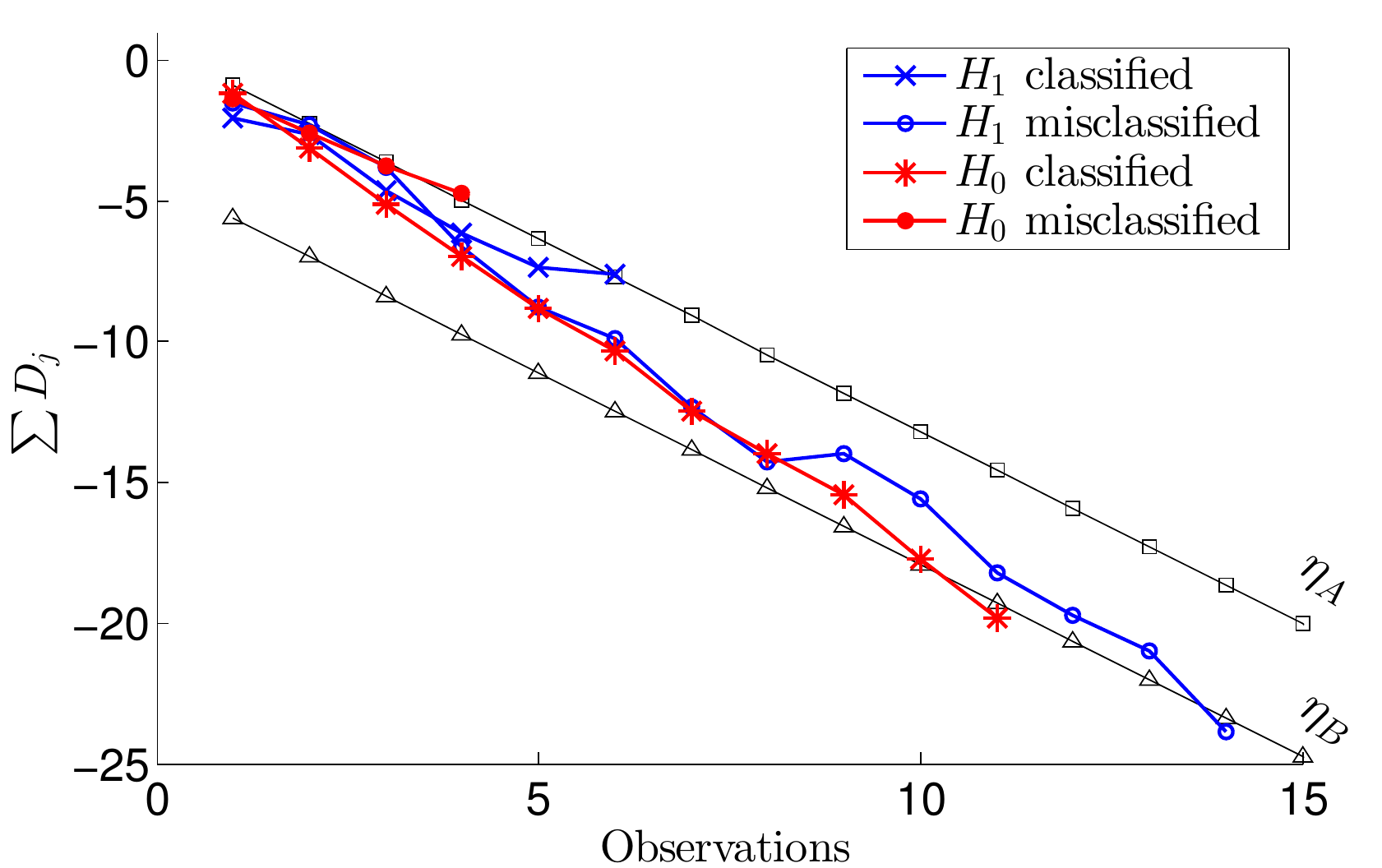}
\caption{SPRT observations track.}
\label{fig:observations}
\end{figure}

For each test image $\bt y_t$, pseudorandom non-overlapping subsets ${\mathcal S}_j$ of size $1024$ pixels, $j=1, \cdots, N$, are taken. The maximum number of observations $N$ is fixed to $256$ because we have experimentally found that for TIFF images a size of $512 \times 512$ should be enough for successful PRNU detection. The results obtained after the SPRT described in \secref{SPRT} are shown in Table~\ref{tab:results}, where $\bar{n}_{H_0}$ and $\bar{n}_{H_1}$ denote the average number of observations that the SPRT needs in order to make a decision for $H_0$ and $H_1$, respectively. The parameters for the hypothesis $H_0$ were set to $\alpha_0 = 1.24$ and $c_0 = 1.78$
after applying the maximum likelihood estimation criterion. These parameters remain fixed throughout all the experiments. Figure \ref{fig:observations} shows the SPRT observation track of some classified and misclassified images.
\begin{table}[!ht] 
%% increase table row spacing, adjust to taste
\renewcommand{\arraystretch}{1.3}
% if using array.sty, it might be a good idea to tweak the value of
% \extrarowheight as needed to properly center the text within the cells
\caption{Experimental results for both SPRTs. $P_D^* = 0.98$ and $P_F^* = 0.3$.}
\label{tab:results}
\centering
%% Some packages, such as MDW tools, offer better commands for making tables
%% than the plain LaTeX2e tabular which is used here.
\scalebox{0.85}{
\begin{tabular}{|l||c|c|c|c||c|c|c|c|}
\hline
\multirow{2}{*}{\bfseries{Device}} & \multicolumn{4}{c||}{\bfseries{Proposed SPRT}} & \multicolumn{4}{c|}{\bfseries{SPRT with fixed $\mu$ and $\sigma^2$}} \\
\cline{2-9}
 & $P_D$ & $\bar{n}_{H_1}$ & $P_F$ & $\bar{n}_{H_0}$ & $P_D$ & $\bar{n}_{H_1}$ & $P_F$ & $\bar{n}_{H_0}$ \\
\hline
Canon 600D &0.9895 & 1.07 & 0.0083 & 1.73 & 0.9895 & 1.08 & 0.0078 & 1.75 \\
Canon 1100D \#1 & 0.9398 & 2.61 & 0.0933 & 4.24 & 0.9323 & 2.50 & 0.0897 & 3.97 \\
Canon 1100D \#2 & 0.9028 & 6.24 & 0.1449 & 10.04 & 0.8750 & 6.53 & 0.1439 & 10.17 \\
Canon 1100D \#3& 0.9880 & 3.98 & 0.1451 & 9.04 & 0.9940 & 4.02 & 0.1450 & 9.25 \\
Nikon D60 & 0.9660 & 6.11 & 0.1869 & 8.65 & 0.9456 & 5.84 & 0.1874 & 9.50 \\
Nikon D90 & 0.8800 & 5.18 & 0.1870 & 10.94 & 0.8800 & 5.50 & 0.1797 & 10.72 \\
Nikon D3000 & 0.9778 & 2.59 & 0.1338 & 4.64 & 0.9722 & 2.67 & 0.1336 & 4.63\\
Nikon D3200 & 0.9900 & 2.63 & 0.1276 & 4.60 & 0.9850 & 2.51 & 0.1298 & 4.58 \\
Nikon D5100 & 0.9750 & 7.86 & 0.2095 & 12.51 & 0.9800 & 8.45 & 0.2162 & 17.57 \\
Nikon D7000 & 0.9350 & 5.45 & 0.2106 & 10.45 & 0.9200 & 5.42 & 0.2021 & 9.27 \\
\hline
\end{tabular}
}
\end{table}

As seen in Table~\ref{tab:results}, for several cameras the empirical values of $P_D$ do not meet the 
target of $0.98$. For this reason, we have applied the correction methods discussed in \subsecref{missmodeling}, obtaining the results reported in Table~\ref{tab:results_2}.

%**** explicar que se toma una decision para casi todas las muestras, excepto algunas, muy pocas en algunas cámaras; poner ejemplo ilustrativo en la grafica de evolucion del test ****

%**** explicar mejor estos resultados y la comparacion entre ambos ****

\begin{table}[!h] 
%% increase table row spacing, adjust to taste
\renewcommand{\arraystretch}{1.3}
% if using array.sty, it might be a good idea to tweak the value of
% \extrarowheight as needed to properly center the text within the cells
\caption{Experimental results for both SPRTs. $P_D^* = 0.98$, $P_F^* = 0.3$, $p=0.0285$, $\beta = 0.65$.}
\label{tab:results_2}
\centering
%% Some packages, such as MDW tools, offer better commands for making tables
%% than the plain LaTeX2e tabular which is used here.
\scalebox{0.85}{
\begin{tabular}{|l||c|c|c|c||c|c|c|c|}
\hline
\multirow{2}{*}{\bfseries{Device}}  & \multicolumn{4}{c||}{\bfseries{Proposed SPRT}} & \multicolumn{4}{c|}{\bfseries{SPRT with fixed $\mu$ and $\sigma^2$}} \\
\cline{2-9}
& $P_D$ & $\bar{n}_{H_1}$ & $P_F$ & $\bar{n}_{H_0}$ & $P_D$ & $\bar{n}_{H_1}$ & $P_F$ & $\bar{n}_{H_0}$ \\
\hline
Canon 600D & 1.0000 & 1.14 & 0.0120 & 3.85 & 1.0000 & 1.14 & 0.0117 & 3.92 \\
Canon 1100D \#1 & 0.9737 & 3.31 & 0.1389 & 10.08 & 0.9774 & 3.30 & 0.1327 & 9.43 \\
Canon 1100D \#2 & 0.9861 & 6.35 & 0.2242 & 24.70 & 0.9722 & 6.58 & 0.2227 & 25.25 \\
Canon 1100D \#3& 1.0000 & 3.44 & 0.2225 & 22.24 & 1.0000 & 3.41 & 0.2223 & 22.80 \\
Nikon D60 & 0.9864 & 6.12 & 0.2715 & 21.12 & 0.9728 & 5.67 & 0.2740 & 23.39 \\
Nikon D90 & 0.9400 & 8.00 & 0.2760 & 26.81 & 0.9300 & 8.69 & 0.2658 & 26.78 \\
Nikon D3000 & 0.9944 & 2.30 & 0.1916 & 11.13 & 0.9944 & 2.32 & 0.1891 & 11.16 \\
Nikon D3200 & 1.0000 & 2.35 & 0.1796 & 11.10 & 1.0000 & 2.32 & 0.1813 & 11.04 \\
Nikon D5100 & 0.9900 & 7.38 & 0.3039 & 30.10 & 0.9900 & 7.14 & 0.3168 & 44.10\\
Nikon D7000 & 0.9900 & 6.48 & 0.3067 & 25.46 & 0.9700 & 6.94 & 0.2923 & 22.55 \\
\hline
\end{tabular}
}
\end{table}

An important analysis over the proposed SPRT is to measure the computational savings achieved with respect to a full-image test. Given $O_F$ as the computational cost of classifying a full image, and $O_S$ the computational cost under our SPRT detector, both are directly proportional to the respective number of pixels. The total number of pixels is $M$, where as for the SPRT detector is $M' = \bar{n} T$, where $\bar{n}$ is the average number of observations which can be written as
\begin{equation}
\bar{n}=\bar{n}_{0} \cdot p_{H_0} + \bar{n}_{1} \cdot p_{H_1}, 
\end{equation}
with $\bar{n}_0$, $\bar{n}_1$ the average of $\bar{n}_{H_0}$, $\bar{n}_{H_1}$ over all devices in Table~\ref{tab:results_2}, and $p_{H_0}, p_{H_1}=(1-p_{H_0})$ the prior probabilities of hypotheses $H_0$, $H_1$, respectively.

On the other hand, when testing a large database with the SPRT, the computational cost is proportional to \mbox{$O_S+[P_D^* p_{H_1} + P_F^* p_{H_0}] \cdot O_F$}, where term in brackets is the probability that the test gives a (true or false) positive. Then, the saving is given by the following ratio
\begin{equation}
\label{eq:Crelation}
\frac{O_S}{O_F} = P_D^* \cdot p_{H_1} + P_F^* \cdot p_{H_0}  + \bar{n} T /M. 
\end{equation}
 
For a database with images of size $M=2000 \times 3000$ pixels and $p_{H_1}=0.01$, a sequential detector with subsets of size $T=1024$, and the $P_D^*, P_F^*$ values of Table~\ref{tab:results_2},  the ratio in \eqref{eq:Crelation}  is approximately $0.3$. Furthermore, the computation of the estimated variance following the simplification in \secref{fve} would produce an additional huge reduction of \mbox{$ 1/M \approx 1.6\cdot 10^{-7}$}.

Finally, in order to compare the proposed detector in \secref{improved} (I-SPRT) and its version with fixed $\mu$ and $\sigma^2$ \mbox{(F-SPRT)}, we averaged over the cameras the respective detection probabilities in Table~\ref{tab:results_2} as well as the average number of observations $\bar{n}$. Notice that we do not compare here on the basis of $P_F$, since the images wrongly classified as $H_1$ will be analyzed a second time (see \figref{Diagram}).
We obtained the following results: $\bar{P}_D^I = 0.986$, $\bar{P}_D^F = 0.981$, $\bar{n}^I = 18.5$ and $\bar{n}^F = 19.9$, where the superscripts $I$ and $F$ denote \mbox{I-SPRT} and \mbox{F-SPRT}, respectively. As we can see, the improved detector offers a small gain in both indicators, so its use is advised despite the extra computation required in the training phase.  

\section{Conclusions}
\label{sec:conclusions}
In this paper we have shown how Wald's sequential test can be implemented for PRNU detection purposes, with the advantage of enabling a very fast test that makes a reliable decision with a minimum number of observations. The test is corrected to account for the event that under $H_1$ the observations may contain very weak fingerprints. The proposed tests are especially useful when very large databases must be searched for device identification.

 \section*{Acknowledgments}
 \scriptsize{Research supported by the Illegal use of Internet (INT) call within the Prevention of and Fight against Crime 
 (ISEC) programme of the Home Affairs Department of the European Commission under project NIFTy (Project Number
 HOME/2012/ISEC/AG/INT/4000003892), the  European Regional Development Fund (ERDF) and the Galician Regional 
 Government under agreement for funding  the Atlantic Research Center for Information and Communication 
 Technologies (AtlantTIC),  the Spanish Government and the European Regional Development Fund (ERDF) under 
 project TACTICA, the European Regional Development Fund (ERDF) and the Spanish Government under project 
 COMONSENS (CONSOLIDER-INGENIO 2010 CSD2008-00010), and the Galician Regional Government under projects 
 "Consolidation of Research Units" 2009/62, 2010/85.}

\bibliographystyle{IEEEtran}

\bibliography{biblio_short}

\end{document}